\input{aipcheck}

\documentclass[
    ,final            
  ]
  {aipproc}

\layoutstyle{6x9}

\usepackage{indentfirst}

\begin{document}

\title{Building complex networks through classical and Bayesian statistics - a comparison}


\author{Lina D. Thomas}{
  address={Instituto de Matemática e Estatística, Universidade de São Paulo, São Paulo, Brazil}
}

\author{Victor Fossaluza}{
  address={Instituto de Matemática e Estatística, Universidade de São Paulo, São Paulo, Brazil}
}

\author{Anatoly Yambartsev}{
  address={Instituto de Matemática e Estatística, Universidade de São Paulo, São Paulo, Brazil}
}

\begin{abstract}

This research is about studying and comparing two different ways of building complex networks. The main goal of our study is to find an effective way to build networks, particularly when we have fewer observations than variables. We construct networks estimating the partial correlation coefficient on Classic Statistics (Inverse Method) and on Bayesian Statistics (Normal - Inverse Wishart conjugate prior). In this current work, in order to solve the problem of having less observations than variables, we propose a new methodology called {\it local} partial correlation, which consists of selecting, for each pair of variables, the other variables most correlated to the pair. We applied these methods on simulated data and compared them through ROC curves. The most attractive result is that, even though it has high computational costs, to use Bayesian inference on trees is better when we have less observations than variables. In other cases, both approaches present satisfactory results.

\end{abstract}

\maketitle


\section{Introduction}

Building networks is a topic that is increasingly present among researchers. They can represent many situations among the most varied subjects. As an example we can mention the Internet, social networks of kinship or other characteristic that connects individuals, organizational networks or relationships between companies, neural networks, food chain, logistics, networks of references among articles.


The main motivation of this work was the article \cite{mine:13}, where regulatory networks were constructed. Data were taken from a case-control study on cervical cancer. In order to build a gene expression network, the authors used the inverse method to calculate the partial correlation coefficient.


However, in studies of {\it microarrays}, it is common to have thousands of genes versus tens of observations, as in Mine et all \cite{mine:13}. This raises the problem of identifiability, where we do not have enough information to make inferences about the data. Statistically, the more knowledge we have about the data, more realistic are the results. To study data with more variables than observations, it is necessary to develop new methodologies.


Although our objective in this study is not to find the best among the existing methodologies, we found that there are different ways to handle this situation. In Peng et all \cite{peng:09} the partial correlation was estimated by linear regression in sparse sets. In the literature, we find other references to applications in data from {\it microarray}. In Han et all \cite{han:08}, the correlation coefficient was calculated on clusters of genes. In the articles Toh et all \cite{toh:01}, Wang et all \cite{wang:03}, Wu et all \cite{wu:03} and Shimamura et all \cite{shimamura:07}, Gaussian graphical models were used in the construction of regulatory networks. In Toh et all \cite{toh:01} cluster analysis was also used, while in Shimamura et all \cite{Shimamura:07} weighted Lasso was used in graphical Gaussian models. The Lasso is a shrinkage and selection method for linear regression. It minimizes the usual sum of squared errors, with a bound on the sum of the absolute values of the coefficients. Gene sets were used in Goeman et all \cite{goeman:07} for obtaining gene expression networks.


\section{Objectives}

The aims of this work, using simulated data, is:

\begin{enumerate}

\item To study and compare the construction of networks using partial correlation coefficients with two methodologies. The first is through classical statistics. For estimation of the partial correlation coefficient, we use the inverse method, as in Mine et all \cite{mine:13}, and for the hypothesis testing, the value calculated by the Fisher's $z$-transform. The second methodology is based on the Bayesian school. We estimate the covariance matrix and use the entries of this matrix to calculate the partial correlation coefficients. We test the hypotheses with the $e$-value from FBST - Full Bayesian Significance Test (see Pereira et all \cite{pereira:99}).


\item To study and compare the construction of networks when we have more variables than observations using what we called local partial correlation. In this method, for each pair of variables, we first select a subset of variables correlated to the pair and then calculate the partial correlation coefficient. This study is also done using the two methods described in the previous item.


\end{enumerate}

\section{Graphical Modeling}
\label{secmodgraf}

We assume that the variables $ X_1, X_2, \cdots, X_p$ follow a density function joint probability 
$$ f_\theta(x_1, x_2, \cdots, x_p), $$ 
\noindent where $ \theta $ symbolizes the unknown parameters. The data are obtained from an 
experiment in which $ n $ observations are independently sampled according to $ f_\theta $.\\


Our goal is to work with models through networks, or graphs, which are finite sets $\mathcal{V}$ of vertices, or nodes, that connect with each other through edges  $\mathcal{E}$. In graphical modeling $\mathcal{V}$ is the set of variables in the model and $\mathcal{E}$ indicates whether there is conditional independence between variables, i.e., for each pair of variables $X_i$ and $ X_j $, there is only one edge between them if the variables are not conditionally independent, given the other variables, $ X_{-ij} $. There are various types of networks, but we will focus on trees.


\section{Conditional Independence}
\label{capindcond}

As we have seen in Section \ref{secmodgraf}, conditional independence is fundamental to graphical modeling. However, despite being based on probability theory, its calculation can be quite complex. Therefore, in practice, it is reasonable to replace conditional independence by zero partial correlation or zero conditional correlation. We decided to work with zero partial correlation, since it is easier to compute than conditional correlation, which depends on the shape of the distribution.


Baba et all \cite{baba:04} show that partial correlation is identical to the conditional correlation if the conditional correlation is independent of the condition and also the conditional expectation is linear. They also prove that the zero partial correlation or the zero conditional correlation does not imply conditional independence, except in the normal distribution case.



\section{Partial Correlation} 

We are interested in the conditional independence of $X_i$, $n \times 1$ and $X_j$, $n \times 1$, given $Y = X_{-ij}$, $i, j = 1, \cdots p , i \neq j$, especially through partial correlation. We can calculate the partial correlation in three different ways that lead to the same results. The first way is by using the original definition based on linear regression. The first requires more hard work, the second method uses the inverse method and the third calculates the partial covariance matrix.


\subsection{Linear Regression}



Let $Y$ be a $k$-dimensional random vector correlated with $X_i$ and $X_j$. If we calculate the correlation coefficient between $ X_i $ and $ X_j$ we will probably conclude that there is some correlation between them. However, we do not know if this correlation is due to the correlation between these two variables and $Y$, or if they really are correlated independent of $Y$. If we keep $ Y $ fixed, does a change in $ X_i $ influence $ X_j$? The main idea of this approach is to remove the effect of $Y$, examining the correlation between the residuals of the projections of $ X_i$ and $ X_j$ in the linear space generated by $Y $, i.e., subtracting the part of the linear relation that comes from $ Y $. Mathematically we have:


$$X_i=\alpha_i+ Y\beta_i+\epsilon_i$$
$$X_j=\alpha_j+ Y\beta_j+\epsilon_j,$$

\noindent where $\beta_i=(\beta_{i1},\cdots,\beta_{ik})^T$ and $\beta_j=(\beta_{j1},\cdots,\beta_{jk})^T$.


Through the least squares linear regression we obtain the following residuals


$$Res_i=X_i-\hat{X}_i$$
$$Res_j=X_j-\hat{X}_j.$$

Note that $Res_i$ and $ Res_j $ are orthogonal to $Y$ and therefore the correlation $\rho(Res_i, Res_j)$ is the correlation between the components of  $X_i$ and $X_j $ that doesn't show linear dependency with $Y$. Therefore, the partial correlation is given by


$$\rho_{ij.Y}=\rho(Res_i,Res_j)=\rho(X_i-\hat{X}_i,X_j-\hat{X}_j)$$


Here $\hat{X}_i=E(X_i)+\Sigma_{X_iY}\Sigma_{YY}^{-1}(Y-E(Y))$ is the projection of $X_i$, i.e., the conditional expectation of $X_i$ given $Y$, where $\Sigma_{X_iY}$ is the conditional matrix of $X_i$ with $Y$ and $\Sigma_{YY}$ is the conditional matrix of $Y$. The construction of $\hat{X}_j(Y)$ is analogous to $\hat{X}_i$\\

$\rho_{ij.Y} \approx 0 \Rightarrow X_i$ e $X_j$ are not correlated if we don't consider $Y$.

\subsection{Inverse Method}


Let  $ R $ be the correlation matrix, where $r_{ij} = \rho(X_i, X_j)$ are the elements of the matrix $R$. If $ R $ is invertible, then we define $P = R ^ {-1} $, where $ p_{ij} $ are the elements of the matrix $P$. The partial correlation is given by:


\begin{equation}\rho_{ij.Y}=-\frac{p_{ij}}{\sqrt{p_{ii}p_{jj}}}\label{eq4}\end{equation}

Note that if the number of observations $n$ is less than the number of random variables $p$, then $ R $ is singular and we cannot use this method.




\subsection{Partial Covariance}
\label{seccovpar}

The partial covariance matrix of $X_{ij}=(X_i,X_j)$ is denoted by

$$\Sigma_{ij.Y}=\left[\begin{array}{cc}
\sigma_{ii.Y}&\sigma_{ij.Y}\\
\sigma_{ji.Y}&\sigma_{jj.Y}
\end{array}\right],$$
which can be calculated through the following


\begin{equation}\Sigma_{ij.Y}=\Sigma_{X_{ij}}-\Sigma_{X_{ij}Y}\Sigma_{Y}^{-1}\Sigma_{YX_{ij}},\label{eq3}\end{equation}

\noindent partitioning the covariance matrix of $(X_{ij})$ in 

$$Cov\left(\left[\begin{array}{cc}
X_{ij}\\Y\end{array}\right]\right) = \left[\begin{array}{cc}
\Sigma_{X_{ij}}&\Sigma_{X_{ij}Y}\\
\Sigma_{YX_{ij}}&\Sigma_{Y}
\end{array}\right],$$ where $\Sigma_{X_{ij}}$ is $2\times 2$, $\Sigma_{X_{ij}Y}$ is $2\times (p-2)$, $\Sigma_{YX_{ij}}$ is $ (p-2)\times 2$ and $\Sigma_{Y}$ is $(p-2)\times (p-2)$. \\


\noindent The partial correlation is given by

\begin{equation}\rho_{ij.Y}=\frac{\sigma_{ij.Y}}{\sqrt{\sigma_{ii.Y}\sigma_{jj.Y}}}.\label{EQ3}\end{equation}\\

Takahashi \cite{takahashi:08} shows the equivalence between the inverse method and the partial correlation method with the original definition (linear regression).


\section{Methodology}

\subsection{Classical Statistical}

The first step was to simulate trees in which the amount of ``children'' follows a lognormal(1,1) distribution. The choice of this distribution was based on Mine et all \cite{mine:13}. But how do we define the relationship of ``parent'' and ``children''? ``Children'' are a linear combination of the ``parent'' with an additional white noise. 

Using simulated data, we estimated the partial correlation using the inverse method, equation (\ref{eq4}).



Since we are looking for non-zero or zero partial correlations to define whether there is an edge between each pair of vertices, we need to test the null hypothesis $ H_0:\rho_{ij.Y}=0$. For this purpose, we can use the Fisher's $z$-transform:


$$z(\hat{\rho}_{ij.Y})=\frac{1}{2}\ln \left(\frac{1+\rho_{ij.Y}}{1-\rho_{ij.Y}}\right).$$\\

We reject $H_0$ with significance level $\alpha$ if:


\begin{equation}
\sqrt{n-|Y|-3}\cdot |z(\rho_{ij.Y})|>\Phi^{-1}(1-\alpha/2),
\end{equation}

\noindent where $\Phi(\cdot)$ is the $Normal(0,1)$ cumulative distribution function and $n$ is the number of observations. For more information see Kendall et all \cite{kendall:73} and \cite{fisher:24}.

\subsection{Bayesian Statistics}

\subsubsection{A Priori Choice}

We assume that the data follows a normal distribution with mean $\mu$ and covariance matrix $\Sigma$. Due to the ease of calculations, we decided to work with the conjugate prior, i.e., $\Sigma \sim IW(k_0,\Psi_0)$ and $\mu \mid \Sigma \sim N(\lambda_0, \frac{1}{v_0} \Sigma)$. \\


However, if we have no information about the parameters, a researcher can use a pilot sample from the same population of the study data in order to make their prior assumptions about the parameters.\\


For this task, we simulate two trees with the same adjacency matrix, linear dependence and number of observations. One is considered the pilot sample and the other is seen as the study data. Denote the pilot matrix by $Z$. \\


We should choose a priori with an average close to what we expect, but with high variance. The Inverse Wishart has expectation,
$$E(\Sigma)=\frac{\Psi_0}{k_0-p-1}, ~~k_0-p>1, $$ 

and variance

\begin{equation} Var(\sigma_{ii})= \frac{2q_{ii}^2}{(k_0-p-1)^2 (k_0-p-3)},~~k_0-p>3,\label{variwii}\end{equation}
\begin{equation} Var(\sigma_{ij})=\frac{q_{ii}q_{jj}+\frac{k_0-p+1}{k_0-p-1}q_{ij}^2}{(k_0-p)(k_0-p-1)(k_0-p-3)}\label{variwij}\end{equation}.\\

Equations (\ref{variwii}) and (\ref{variwij}) suggest that the lower $k_0$, the greater the variance. And since the existance of $E(\Sigma)$ requires that $k_0> p +1$, we chose $k_0 = p +3$. Based on $E(\Sigma)$, we took $\Psi_0 = (k_0-p-1)Cov(Z)$, where $Cov(Z)$ is the maximum likelihood estimator of the covariance of the pilot sample. Then we used the partial covariance (\ref{EQ3}) for the calculation of partial correlation.\\



To test the null hypothesis of zero partial correlation, we used FBST. Note that we need the distribution of partial correlation. But it is only possible to find the distribution of the partial covariance, which also follows an Inverse Wishart distribution (see Mardia et all \cite{mardia:79}).\\


To enable the calculation of the $e$-value, we simulated the partial covariance $1000$ times and estimated the density of the partial correlation with the function {\it density} of R. To calculate the area of such density that will provide the $e$-value, we use the approximation of the area by rectangles.


\section{Simulation} \label{capsim}

In this part of the work, to confirm the effectiveness of the theory described above and find the best method, we simulated 500 trees with 100 variables with the following algorithms:

\begin{enumerate}
\item The tree starts with one vertex in the first generation.
\item Generate the number $n_1$ of offsprings of $X_1$ from a Lognormal$(1,1)$ distribution, with the condition that the first vertex must have at least one descendent.
\item For each vertex of the second generation, $(X_2,\cdots,X_{1+n_1})$, generate the number of offsprings $n_{2i}$, $i=1,\cdots,n_1$, from a Lognormal(1,1) distribution, with the condition that the second generation must have at least one descendent and $1+n_1+n_2\leq p$, $n_2=n_{21}+\cdots+n_{2n_1}$.
\item For each following generation, repeat the last item until we get $p$ variables.
\end{enumerate}

Now that we have all the parental structure, we need to simulate the data.

\begin{enumerate}
\item Generate the first variabel $X_1$ from a Normal(0,1) distribution
\item The relation between parents and offsprings comes from a linear combination. We generate the $p-1$ coefficients of the $p-1$ linear combinations from a Uniform$(0,7;2)$ distribution. The positive and negative signals of the coefficients follow a Bernoulli$(\frac{2}{3})$ distribution, where success is the positive signal.
\end{enumerate}

\label{alg1}

We calculated the partial correlation, the p-value and e-value. In case we have more variables than observations, we computed the local partial correlation. To ascertain the quality of the model, we drew the ROC curve in classical and Bayesian approach for each tree, calculated the average Bayesian and classical ROC curves' area and then compared them. This was done for different numbers of observations: $ 50, 250, 500$ and $1000$.


\subsection{Local Partial Correlation}

\begin{figure}[h]
\includegraphics[height=7cm]{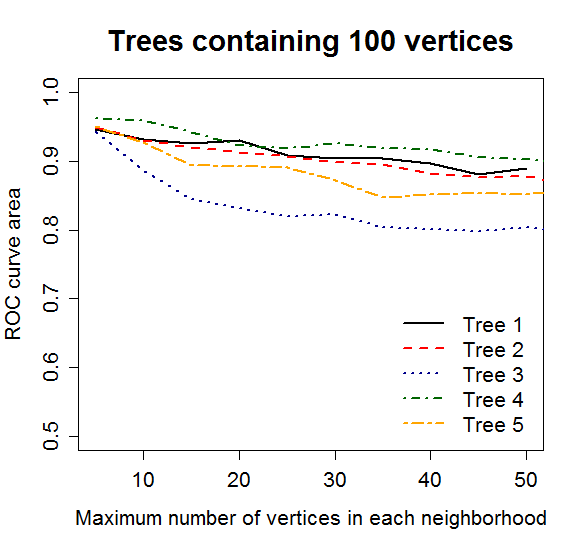}
\caption{Maximum number of variables to calculate the partial correlation of 5 simulated trees.}
\label{graficocpl}
\end{figure}

In this calculation, for each pair of significantly correlated variables, we began by considering only the vertices correlated to the the pair, which we called neighborhood. If this neighborhood contains more vertices than observations, we select the $\frac{n}{10}$ vertices with the highest correlations. These values were selected according to the number of vertices that generated the greatest area under the ROC curve as we can see in Figure \ref{graficocpl}.


\subsection{ROC curves analysis}

\begin{figure}[h]
\includegraphics[height=6cm]{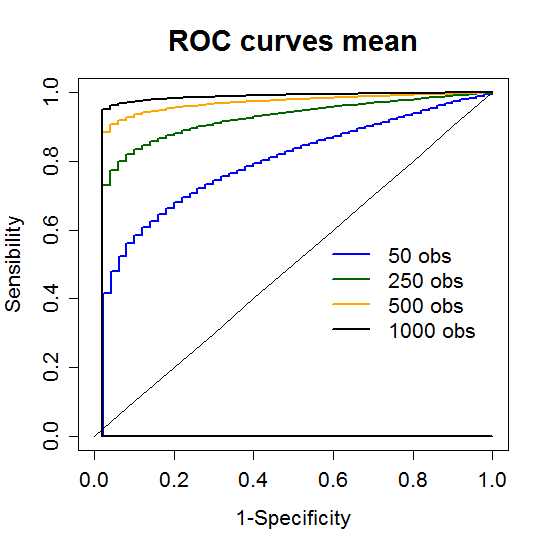}
\includegraphics[height=6cm]{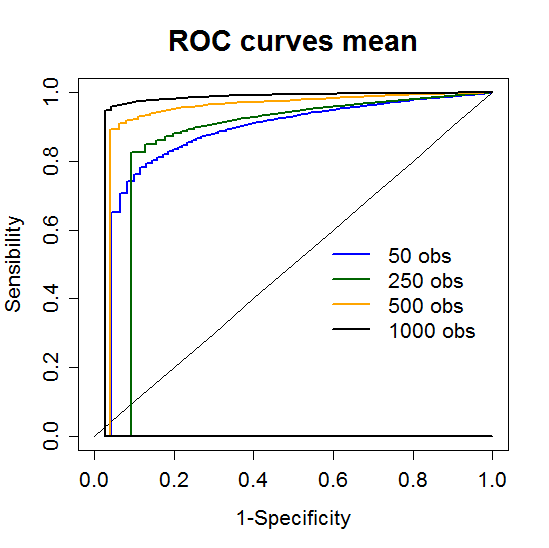}
\caption{Comparing the means of the ROC curves for 50, 250, 500 and 1000 observations through classical (left) and Bayesian (right) statistics}
\label{grafico1}
\end{figure}

Observing Figure \ref{grafico1}, we can see that the area of the curves increase with the number of observations, as expected. The most interesting remark is that the area of the curve with $50$ observations is pretty close to the area of the curve with $250$ observations, especially in the Bayesian approach. Since we are working with networks composed of $100$ variables, we conclude that the ROC curves of the trees with less observations than parameters ($n=50,p=100$) are relatively close to the ROC curves of the trees with more observations than parameters ($n=250,p=100$).


\subsubsection{Comparing the Bayesian and Classical ROC curves}

\begin{figure}
\includegraphics[height=7cm]{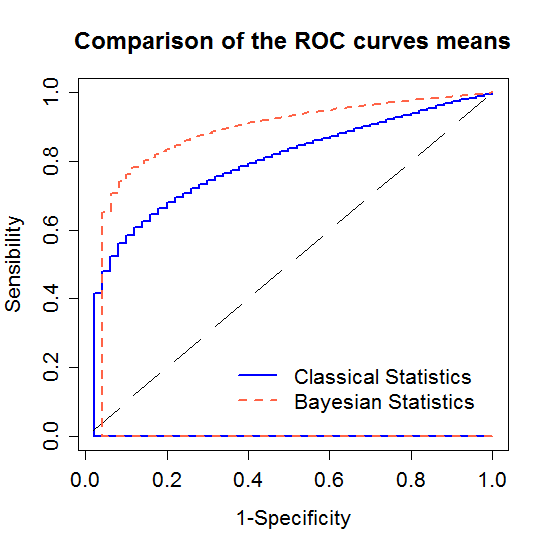}
\includegraphics[height=7cm]{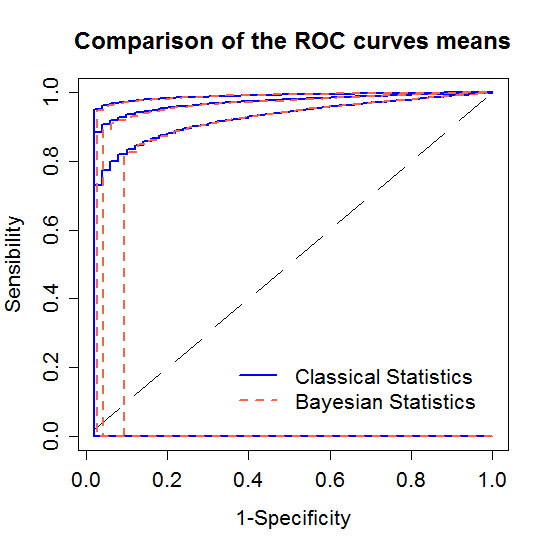}
\caption{Comparison between the ROC curves means with classical and Bayesian statistics for 50 (left), 250, 500 e 1000 obs (right)}
\label{grafico7}
\end{figure}

Comparing the Bayesian and Classical ROC curves, as seen in Figure \ref{grafico7} (right), when we have more observations $(250, 500, 1000)$ than variables $(100)$, the mean of the ROC curves in the classical and Bayesian approaches are superimposed, indicating that one is as good as the other. In the other hand, the ROC curve mean in the Bayesian approach, Figure \ref{grafico7} (left), with only $50$ observations shows a larger area than the classical ROC curve mean. Although the computational cost is greater in the Bayesian statistics, its use is advisable in cases when we have more variables than observations.

\bibliographystyle{aipproc}   

\end{document}